# Measurement and analysis of the Doppler broadened energy spectra of annihilation gamma radiation originating from clean and adsorbate-covered surfaces.

S. Lotfimarangloo[a], V. A. Chirayath[a,*], P. A. Sterne[b], H. Mahdy[a], R. W. Gladen[a,c], J. Driscoll[a], M. Rooks[a], M. Chrysler[a], A. R. Koymen[a], J. Asaadi[a], A. H. Weiss[a,+]

[a] Department of Physics, University of Texas at Arlington, Texas, USA 76019
[b] Lawrence Livermore National Laboratory, Livermore, CA 94550, USA.
[c] Department of Physics, The University of Tokyo, Tokyo 113-0033, Japan.
[*]chirayat@uta.edu, [+]weiss@uta.edu

## Abstract

We present measurements and theoretical modeling demonstrating the capability of Doppler Broadened annihilation gamma Spectroscopy (DBS) to provide element-specific information from the topmost atomic layer of surfaces that are either clean or covered with adsorbates or thin films. Our measurements show that the energy spectra of Doppler-shifted annihilation gamma photons emitted following the annihilation of positrons from the topmost atomic layers of clean gold (Au) and copper (Cu) differ significantly. With the aid of the positron annihilation-induced Auger electron spectroscopy (PAES) performed simultaneously with DBS, we show that measurable differences between the Doppler broadened gamma spectra from Au and Cu surfaces in the high energy region of the gamma spectra can be used for the quantification of surface chemical composition. Modeling the measured Doppler spectra from clean Au and Cu surfaces using gamma spectra obtained from ab initio calculations after considering the detector energy resolution and surface positronium formation pointed to an increase in the relative contribution of gamma from positron annihilations with valence shell electrons. The fit result also suggests that the surface-trapped positrons predominantly annihilated with the delocalized valence shell (s and p) electrons that extended into the vacuum as compared to the highly localized d electrons. Simultaneous DBS and PAES measurements from adsorbate (sulfur, oxygen, carbon) or thin film (selenium (Se), graphene) covered Cu surface showed that it is possible to distinguish and quantify the surface adsorbate and thin-film composition just based on DBS. DBS of elemental surfaces presents a promising avenue for developing a characterization tool that can be used to probe external and internal surfaces that are inaccessible by conventional surface science techniques.

## 1. Introduction

Data obtained using PAES has provided clear evidence that the electron signals originating following the annihilation of surface-trapped positrons originate almost entirely from the topmost atomic layer of the sample surface [1, 2]. The enhanced surface selectivity arises from the limited overlap of the surface-trapped positron wavefunction with the electron wavefunctions beyond the first atomic layer [2, 3]. Since the PAES signal originates following the Auger decay of annihilation-induced core holes, a strong correlation can be expected between the PAES signal and the energy of the Doppler-shifted gamma photons that originate following the annihilation of surface-trapped positrons with core and valence electrons. Therefore, it can be reasonably expected that the annihilation gamma spectrum measured using DBS may reflect the elemental composition of the topmost atomic layer with a sensitivity similar to PAES.

The ability of Doppler broadened annihilation gamma spectroscopy to probe the chemical environment at the site of positron annihilation has been employed widely for examining the chemical environment of open-volume defects inside materials [4,5,6,7]. Specifically, Coincidence Doppler broadening (CDBS) measurements from the bulk of a range of pure elements demonstrated the feasibility of differentiating between elements [8, 9]. The CDBS measurements have revealed that in addition to the amount of broadening, the shape of the gamma spectrum reflects the momentum distribution of the core electrons involved in the positron annihilation process. Several studies have shown that it is possible to predict the atomic concentration at the site of the positron annihilation by comparing the Doppler broadened gamma spectra to spectra from elemental samples [6,7]. Angular correlation of the two 511 keV annihilation gamma, which similarly reflects the momentum of the electron-positron pair, has also been employed to identify the surface termination of ligand-capped semiconductor quantum dots encapsulated in thin films [10,11]. However, DBS or CDBS has not been applied for quantitatively probing the chemical environment of external surfaces. A few investigations using a variable energy positron beam applied to investigate the native oxide on Si showed that the Doppler spectra from the Si surface resemble the Doppler spectra from the bulk of quartz [12], providing the first hints that the annihilation gamma spectrum measured from the surfaces can be used for elemental characterization. However, these investigations lacked complementary data showing the surface composition probed by positrons.

The fact that CDBS from the bulk of the sample can be used to probe the elemental nature at the positron annihilation site does not trivially extend to surfaces. Evidence for the complexities in interpreting annihilation data introduced by surface-related phenomena can be clearly seen in the two-dimensional angular correlation measurements on clean low-index aluminum surfaces and aluminum surfaces with oxygen adsorption reported in Chen et al. [13]. In this study, the two-dimensional momentum distribution of the annihilating electron-positron pair measured from the surface was considerably narrower than those measured from the bulk. This was hypothesized as the effect of inhomogeneous electron density at the surface compared to the bulk, the confinement of positrons in a direction perpendicular to the surface, and the contribution from positronium annihilation. It has been challenging to investigate whether the core contributions to the high momentum region of the annihilation gamma spectra from surfaces are significant enough to provide the contrast needed for CDBS to be used as a chemical characterization tool. This is an important question to answer since it can be expected that the most chemical-specific information will be contained in the high momentum region associated with the chemically distinct core levels. The positron wavefunction for the image-potential surface state can be expected to have reduced overlap with core electrons compared to bulk state positrons since the positron is localized in the vacuum outside the surface and has appreciable overlap only with delocalized valence electrons that extend into the vacuum side. A second factor that could interfere with using CDBS data from the surface to perform chemical identification is that a significant fraction of the gamma signals due to positrons at or near the surface may be due to the self-annihilation of positronium (Ps) formed at the surface. The experiments reported in this paper constitute the first measurement of CDBS obtained from surfaces whose elemental content at the site of positron annihilation was measured using the spectra of Auger electrons ejected from the top surface following the decay of the annihilation-induced holes. We believe our data provide definitive answers to the questions posed above regarding the suitability of CDBS for surface studies.

This study utilized the low-energy positron beam system at UTA [14] to perform PAES and CDBS simultaneously from the same surface. This has helped us overcome many difficulties previously encountered in quantitatively discerning surface Doppler spectra. The CDBS measurements were correlated with simultaneous PAES measurements to provide evidence that gamma photons, emitted after the annihilation of surface-trapped positrons, provide surface-selective chemical information. Our analysis of the shape of the annihilation gamma spectrum, obtained from clean and adsorbate-covered surfaces, has shown that the high energy region (corresponding to the high momentum of the annihilating electron-positron pair) of the annihilation gamma reflects the elemental composition, even though the annihilation gamma from core electrons is reduced, and even with the presence of positronium. We can deduce the surface chemical information and orbital annihilation rates by fitting the measured Doppler spectra with ab initio gamma spectra corresponding to positron annihilation from individual orbitals.

Our measurements provide strong support for efforts to develop a new technique based on using CDBS for the chemical characterization of internal surfaces of nanoporous materials. Since the 511 keV annihilation gamma can exit through many millimeters of sample or reaction cell wall without any loss of information, such a method can be used for in-operando characterization of surface composition variations brought about by catalytic reactions or by surface migration driven by reactive gas exposure in nanoporous materials.

The ability to investigate and understand the chemistry occurring within internal surfaces can significantly impact numerous scientific fields, including catalysis, sensing, biomedicine, gas separation, energy storage, etc. [15,16]. Development of such a technique needs calibration spectra from controlled external surfaces, as presented here. In the subsequent sections, we present the experimental and theoretical methods employed to obtain CDBS and PAES from the external surfaces of select samples. We will discuss the implications of the results and possible future work.

## 2. Methods

### i. Experiment

Except for the data shown in Figures 4 and 7, measurements were obtained using a low-energy positron beam system described previously in [14]. Monoenergetic positrons with a mean energy of ~16 eV are magnetically transported through a 1m field-free ToF tube to reach the sample. By adjusting the potential difference between the sample and the ToF tube, the incident kinetic energy of the positrons on the sample can be varied. The PAES and CDBS measurements were carried out with a sample bias of -5 V and ToF tube bias of -25 V. During the energy calibration measurements of PAES, the sample bias was varied from -35 V to - 900 V. At low positron energies, positron annihilation occurs predominantly from the surface state or after the formation of the Positronium atom. A small fraction of surface state annihilations results in the emission of Auger electrons, which, after traveling through the 1m field-free tube, are directed toward a microchannel plate (MCP) electron detector by an **E**x**B** system. The time difference between the detection of the annihilation gamma signal by a NaI scintillation detector placed 2.5 cm away from the sample and the electron detection signal by the MCP is utilized to construct a histogram that represents the measured flight times of the electrons. The as-measured time of flight is converted to the kinetic energy of the electrons using the energy calibration measurements described in [14]. Please note that because of the -20V potential difference between the ToF tube and the sample, electrons with energies lower than 20 eV never make it to the MCP. Therefore, low-energy secondary electrons are removed from our PAES data.

We have recently enhanced the capabilities of our low-energy positron beam system by adding a high-purity germanium (HPGe) detector. This addition has enabled us to simultaneously measure the energy of both annihilation-induced Auger signals and Doppler-shifted annihilation gamma photons from the same surface. The HPGe detector detects the second 511 keV annihilation gamma, which moves antiparallel to the one detected at the NaI from 10 cm away from the sample. We have used conventional NIM logic to ensure a coincidence between the HPGe and NaI scintillation detection signals. The signal from the NaI scintillation detector is amplified and fed to a timing single channel analyzer (TSCA), producing a pulse of 5 µs width. We have used a pulse stretcher built using a monostable multivibrator (IC 74121) to stretch the timing pulse to 35 µs. The time delay between the stretched pulse and the 25 µs Gaussian-shaped HPGe detector pulse from the spectroscopy amplifier is adjusted to ensure overlap. The NaI pulse is input as a gate pulse to the 8K ORTEC Easy MCA that collects the HPGe signals. By enforcing the coincidence condition, we have been able to increase the signal-to-background ratio from 30:1 to 3000:1. The HPGe detector installed at the positron beam system with 1 m ToF has an energy resolution of 1.35 keV at 511 keV peak, which we have determined through energy calibration measurements using both [133]Ba and [152]Eu calibration sources during each experiment. The measured energy resolution is sufficient for experiments that characterize shape variations in the gamma spectrum's high energy region (> 3 keV). The gain was adjusted such that the energy per channel was ~120 eV per channel.

In addition to the measurements using the 1 m ToF system described above, we used the UTA 3m ToF system , described in detail in [17] to measure the CDBS from single-layer graphene on polycrystalline Cu and from the bulk of the surface sputtered polycrystalline Cu. The coincidence condition was set up as described above for the 1m system. The ToF tube was grounded, and the incident kinetic energy of the positron beam ranged from 2 eV to 20 keV. The energy resolution of the HPGe detector installed in this beamline is 1.1 keV at 511 keV.

CDBS and PAES data were collected from the clean, adsorbate, and thin film-covered surfaces of Cu and the clean surface of Au. As received, polycrystalline foils of Au and Cu were inserted into the vacuum chamber. During the thermal outgassing of the ultra-high vacuum (UHV) chamber, the sample temperature reached ~358 K. All measurements were performed with the chamber operating at a pressure at or lower than $10^{-10}$ torr, ensuring that the sample surface remained unchanged during data collection. PAES and CDBS spectra were collected from the polycrystalline Cu surface before cleaning it (the data collected under these conditions are labeled Cu surface – as inserted). Clean surfaces of Au and Cu were obtained by sputtering the sample using an Argon ion sputter gun for ~ 30 minutes at $7x10^{-5}$ torr. Samples were sputtered periodically to ensure that Auger peaks of oxygen or carbon did not appear in the PAES spectra. The PAES and CDBS data collected under these conditions are labeled clean Au or Cu surface. The Cu surface was later oxidized by exposing a newly sputtered Cu surface to an oxygen partial pressure of ~ $1x10^{-4}$ torr while maintaining the sample at 673 K. The oxygen exposure aimed to achieve at least an exposure equivalent to ~ $10^6$ Langmuir. These data are labeled as CuO. Single-layer graphene (SLG) grown on the polycrystalline Cu substrate was purchased from ACS materials. The presence of SLG was independently verified by Raman spectroscopy. Data from these samples are labeled as SLG-Cu. After removing the SLG layer by ion sputtering, we measured the Doppler spectra from the bulk of the polycrystalline sample by increasing the incident positron energy to 20 keV (data labeled as Cu Bulk). We grew a thick film of amorphous Se (~ 1 μm) using thermal deposition on the polycrystalline Cu substrate. The presence of an amorphous layer of Se was confirmed by Raman spectroscopy and X-ray diffraction studies. PAES and CDBS measurements were carried out in the as-deposited condition (labeled as aSe-as received) and after sputtering the sample for ten minutes (aSe-sputter) under similar conditions as described previously.

## ii.     Analysis of the Doppler broadened spectra.

The Doppler broadened annihilation gamma spectrum from each sample was collected such that there were at least 200K counts under the 511 keV peak. Due to the reduced strength of the positron source, this meant that the data was collected over several days. The surface condition was constantly monitored using PAES. Data was saved every two hours to correct any gain shift during the experiment and was later added together after correcting for any channel shift. The raw data were fit directly without any further modifications. The fitting methodology is described for each sample, where their results are discussed.

### iii.    Theoretical Modeling of the Doppler broadened spectra.

To calculate the Doppler spectra from various elements such as Cu, Au, carbon, oxygen, sulfur, and Se, we used an atomic calculation method as described in [18]. This approach doesn't require specific atomic positions, which makes it suitable for our needs, where we are most interested in the shape of the Doppler spectra at higher momentum where the spectrum contains contribution primarily from positron annihilation with inner shell electron states that have an atomic-like character. A standard self-consistent-field atomic program was used [19] to calculate the electronic orbitals by choosing an appropriate electron configuration for each element. The positron wavefunction was obtained by solving single particle Schrodinger equation containing the positron potentials obtained from the self-consistent electron charge density given by the atomic program and the positron-electron correlation potential expressed in terms of $r_s$ [18]. Here, $r_s$ is the radius of the sphere containing one electron, and its value depends on the calculated electron density. Finally, we treated the positron-induced local electron density enhancement, expressed by the enhancement factor $\Gamma(r_s)$ using the generalized gradient approximation given by Barbiellini et al. Some modifications were made (to be described in a future publication) to details of previous calculations made for bulk systems with the goal of taking into account the difference between the wavefunctions of positrons localized in the bulk as opposed to the wavefunctions of positrons trapped in a surface state.

Sterne et al. showed that the Doppler spectra obtained using atomic calculations qualitatively match the shape of the experimental spectrum when annihilation occurs in bulk.  However, the data reported in this work suggests that the theoretical spectrum calculated for the bulk failed to capture the critical regions of the experimental line shape specifically for systems with d-electrons, as simple atomistic calculations cannot accurately provide individual orbital annihilation rates [18]. In the modeling used in this paper we used a linear superposition of the line shapes for individual core levels obtained from the atomistic calculations where the relative contributions to the spectra from each core level were taken as fitting parameters that were chosen with the goal of getting an improved visual fit.  The parameter values obtained from this procedure were then used to estimate the relative annihilation probabilities (details of the fitting procedure applied to each spectrum will be described in respective sections).  The use of a linear combination of calculated core line shapes was motivated by the observation that the atomic-based calculations for individual core levels appeared to provide a reasonable fit to PAES-Gamma coincidence measurements of the Doppler broadened annihilation gamma spectra emitted following the annihilation of positrons with 3p electrons in Cu [21, 22], as well as the  1s electrons in C and 1s electrons in oxygen, and 2p electrons in oxygen [23].

### 3.  Results and Discussion

### i.     Clean Cu surface

The PAES data obtained from the sputtered surface of the Cu foil showed that only the Cu $M_{2,3}VV$ Auger peak and the $M_1VV$ Auger peaks were present (Fig. 2(a)). Therefore, the CDBS data represents the annihilation gamma spectra from a sputtered clean Cu surface. The ratio of the intensity of the $M_{2,3}VV$ peak to the intensity of the $M_1VV$ Auger peak provides the ratio of their orbital annihilation rates since the electron transport and detection are uniform in these energy ranges. From the intensity of the Auger peaks, we can see that the annihilation of positrons with 3p electrons is about 13 times more probable than the annihilation with 3s electrons.

We fit the raw CDBS data from the clean Cu surface as a linear combination of the orbital Doppler broadened annihilation gamma spectra from atomic calculation after convoluting with the instrument resolution function (a Gaussian with an FWHM corresponding to the measured resolution of the detector during that experiment) and normalizing to the total number of counts in the experiment (Fig. 2(b)). The coefficients were determined using the least squares method. The determined parameters include the constant background. The fit gives the contribution of each atomic orbital annihilation to the total Doppler spectra, which is given in Table 1. The annihilation probabilities obtained from the fit are compared to those obtained directly from theoretical calculation, which assumed a $3d^{10}4s^1$ configuration and a $3d^94s^2$ configuration for the Cu atom. We included the contribution of the surface positronium annihilation either in the singlet state or through the pick-off annihilation of the triplet state in our fits. The contribution of the positronium (singlet and triplet) was obtained by analyzing the positron-induced secondary electron spectra using methods developed by [24]. We added a Gaussian with a FWHM equal to the instrument resolution for the singlet annihilation. We used the calculated Doppler spectra from the carbon surface for the pick-off annihilation of triplet or ortho-positronium (o-Ps) from the UHV chamber's walls. This is reasonable as previous investigations have shown that the annihilation gamma spectra produced by the pick-off annihilation of o-Ps represent the surface elemental composition [25]. The contribution from the positronium annihilation was fixed at the value obtained from the analysis of the secondary electron spectra from the clean Cu surface during the least squares fit of the gamma spectrum. During the fit, only the coefficients of 4s, 3d, and 3p orbitals were left free, and the coefficients of the rest of the deep core levels were fixed to what was obtained from the atomic calculations. After the fit, the coefficient of the deep core was adjusted such that the total of all coefficients added up to one.

The contribution of each orbital annihilation to the total annihilation spectrum determined using this method (listed in Table 1) points to the following factors. The annihilation with valence electrons (4s+3d) contributes to 94% of the Doppler spectra from the Cu atom. The ratio between the 3p and 3s orbital annihilation rates is about 15, consistent with what was independently obtained from PAES. The probability of positron annihilation with 3d electrons is less than a tenth of the annihilation probability with 4s electrons, even though there are far more electrons in the 3d orbital (10 or 9) compared to one or two electrons in the 4s shell. The significant suppression of 3d annihilations, as deduced from the fits, is likely due to the reduced overlap of the positron wavefunction with the localized 3d electrons at the surface compared to the highly delocalized 4s electrons. Suppression of the role of 3d electrons in other surface-specific phenomena has been seen previously, like in the neutralization of He ions on Cu surfaces where the contribution of 3d electrons is highly suppressed, and 4s electrons play a prominent role [26]. Recently, Fairchild et al. [27] showed a similar effect in another positron-based surface spectroscopy named Auger-mediated positron sticking-induced electron spectroscopy (AMPSIES). In AMPSIES, spectra of electrons emitted following positron sticking directly reflect the local electron density at the surface. It was shown that the valence band density of states probed by the positron as it approaches the surface differs from what is probed in photoemission spectroscopy. The difference was due to the reduced Coulombic interaction of positrons with the highly localized 3d electrons. A self-consistent calculation of the positron wavefunction at the surface, the state-dependent enhancement factors, and the overlap integral of the positron and electron wavefunctions should be able to bring out the experimentally observed suppression of 3d annihilation at the surface.

## ii.    Clean Au surface

The PAES data obtained from the surface of sputtered Au confirms that positrons only annihilate with electrons of Au (Fig. 3(a)). The PAES data shows three Auger peaks, which correspond to the annihilation of positrons with $5p_{3/2}$ ($O_3VV$ at 45eV), $5p_{1/2}$ ($O_2VV$ at 57eV), $4f_{5/2}$ and $4f_{7/2}$ ($N_{6,7}VV$) electrons. We estimated the ratio of annihilation rates of 5p electrons to 4f electrons by analyzing the Auger peak intensities, which came out to be around 7. Table 2 lists the contribution of the orbital annihilations to the total Doppler spectra obtained after the least squares fit using the Doppler line shapes obtained from atomic calculations (Fig. 3(b)). The positronium component has been determined using the analysis of the secondary electron spectrum. The 5p to 4f annihilation rate ratio obtained from the least squares fit is consistent with the independent estimation of the ratio of annihilation rates from PAES. Just like in the case of a clean Cu surface, our results show a significant reduction in the contribution from 5d electron annihilation to the total Doppler spectra. Specifically, our findings reveal that the contribution from 6s annihilations is seven times greater than the annihilation from 5d electrons, even though there are ten times more electrons in the 5d orbital. Our fits also show that valence (6s+5d) annihilations alone account for 96% of the intensity of the Doppler spectra from the Au surface.

The ratio of the CDBS data from the clean Au surface to that obtained from the clean Cu surface is shown in Fig. 3 (c). The ratio of the fitted theoretical Doppler spectrum of Au to that of Cu is the solid line through the data points. The excellent fit shows that the modified theoretical spectrum obtained through a free fit of the individual orbital line shapes can capture most of the details of the ratio curve. From here on, we will take the CDBS data from the clean Cu surface as our standard; all other spectra will be compared to the clean Cu surface data. The ratio of Doppler spectra from the clean Au surface to that obtained from the clean Cu surface shows that Doppler broadening spectra from surfaces can provide elemental information even though surface Doppler spectra from Au and Cu have (i) different contributions from surface positronium annihilations,

(ii) have a significant fraction of the total annihilation spectrum coming from valence electron annihilations and (iii) show significant reduction in d-electron annihilations even though valence density is dominated by 3d or 5d electrons in Cu or Au respectively.

### iii.    Single Layer Graphene (SLG) on polycrystalline Cu (SLG-Cu)

Chirayath et al. [2] showed that the PAES from SLG contained carbon and oxygen through their respective KVV Auger peaks and Cu through the $M_{2,3}$VV Auger peak. This study estimated that approximately 8% of the annihilation happens at the Cu-SLG interface due to the appreciable positron probability density at the interface, even when the positron is trapped at the surface state of graphene. The surface Doppler spectrum of the SLG was measured using the positron beam with a 3m ToF spectrometer. The average energy of the positrons was about 2 eV. The measured surface Doppler spectrum from SLG (Fig. 4(a)) was fit with the theoretical Doppler spectrum of carbon, oxygen and the model surface Doppler spectra of clean Cu obtained after fitting the measured Doppler spectrum from clean Cu surface (section 3. (i)). The contribution from the singlet Ps annihilation was also kept as a free parameter during the least squares fit since secondary electron spectra were not collected concurrently in this beamline. We obtained reasonable fits to the spectrum without modifying the calculated annihilation probabilities of the individual core levels of carbon and oxygen. The annihilation probability with Cu was found to be ~ 13.5%, close to what was predicted by the PAES [2]. Annihilation probabilities from other elements are given in Table 3. The ratio of the Doppler spectra from SLG on Cu to the clean Cu surface (Fig. 4(b)) shows the ability of the Doppler spectra to distinguish between the clean Cu surface and the clean Cu surface with a single atomic layer of carbon. The solid line through the data points is the ratio of respective fitted curves. The ratio curve and the corresponding theoretical fits show that the Doppler broadening spectroscopy has selectivity to the topmost atomic layer similar to PAES and that it can be effectively used for estimating the percentage of positron annihilation with various elements on the surface, providing more confidence towards its application for the assay of the chemical composition of external and internal surfaces. Please note that while fitting the SLG Doppler spectrum, we did not consider o-Ps pick-off annihilation separately, as the wall annihilations were treated as annihilations from carbon. As far as we know, these are the first attempts at fitting the surface Doppler data as a linear combination of Doppler spectra from multiple elements. Even for bulk, only two-element linear combinations have been used to obtain quantitative estimates of annihilation probabilities with different elements [6,7].

### iv.    Cu surface – As inserted.

The PAES and Doppler spectra obtained from as-inserted polycrystalline Cu foil are given in Fig. 5(a) and 5(b), respectively.   The PAES spectra for this surface shows the presence of the $L_{2,3}$VV Auger peak of sulfur in addition to the $M_{2,3}$VV Cu peak.   The presence of S at the surface is reasonable given that the sample temperature can reach as high as 358K during the thermal outgassing of the UHV system.  Temperatures in this range are known to cause sulfur, a common contaminant in Cu, to diffuse to the surface. An analysis of the PAES spectra showed that ~ 11% of the total annihilations detected correspond to the annihilation of Ps from the singlet state, and ~3% of the annihilations are o-Ps from the chamber's walls.

The Doppler spectrum from the as-inserted Cu surface (5 (b)) was fit with the model Doppler spectra representing the Cu surface (section 3 (i)) and the Doppler spectrum from sulfur obtained from atomic calculations. We used the calculated Doppler spectra from sulfur without changing the orbital annihilation rates. As before, the Ps components were fixed during the least squares fit, and only the coefficients of Cu and sulfur were determined from the fits. Our fits show that annihilations from Cu (49%) are 1.3 times the annihilation probability from sulfur atoms (37%), which is consistent with the analysis of the PAES spectra after considering the annihilation probabilities from sulfur L shell and Cu M shell previously determined by Jensen and Weiss [28]. Fig. 5 (c) shows the ratio of the measured curves (Cu as inserted/clean Cu), indicating that CDBS can capture sulfur segregation on the Cu surface.

**v.    CuO**

The PAES data obtained from CuO (Fig. 6 (a)) reveals the presence of oxygen through the sharp Auger peak at 503 eV, which corresponds to the Auger decay of the annihilation-induced K shell electrons in oxygen. Additionally, there is the Cu $M_{2,3}$VV Auger peak. The absence of the $L_{2,3}$VV sulfur peak indicates that we have effectively sputter-cleaned surface-segregated sulfur atoms and that the surface consists only of Cu and oxygen atoms. Considering the Auger peak intensities and the annihilation probabilities for oxygen 1s and Cu 3p shells as calculated by Jensen and Weiss [28], we estimate that the annihilation with Cu atoms is 2.7 times more likely than annihilation with oxygen atoms. To model the Doppler spectrum from the oxidized surface of Cu (Fig. 6(b)), we used the model Cu Doppler spectra obtained in section 3(i) and the Doppler spectrum from oxygen obtained through atomic calculations. We used the theoretical Doppler spectrum from oxygen without modifying the individual orbital annihilation rates. Our fits indicate that 63% of the positron annihilations are with Cu electrons, and 20% of annihilations are with O electrons, which is consistent with the PAES measurements. The ratio of the measured Doppler spectrum from the CuO surface with the Doppler spectrum from the clean Cu surface (Fig. 6(c)) shows that CDBS can follow chemical composition variation at the surface, like oxide formation. The ratio of the fitted curve matches the overall shape of the ratio of the measured data. However, the fit cannot effectively capture all the details of the experimental curves since the calculated Doppler spectra are for individual elements and not for the oxide film that has formed on the surface with thermal treatment and oxygen exposure.

**vi.    Cu Bulk**

The Doppler broadened spectrum (Fig. 7 (a)) from the bulk of the sputter-cleaned polycrystalline Cu was collected using the positron beam with the 3m ToF and at an incident positron beam energy of 20 keV. The measured Doppler spectrum was fit using the theoretical

Doppler spectrum of carbon, sulfur, and bulk Cu. Carbon and sulfur were included as common contaminants in polycrystalline Cu. During the least squares fit, the annihilation probabilities of the individual Cu orbitals 4s, 3d, and 3p were allowed to vary. However, the annihilation probabilities of the rest of the orbitals (3s, 2p, 2s, 1s) were kept constant. The contribution of carbon and sulfur was also allowed to vary. After the minimization of the reduced $\chi^2$, the coefficients or annihilation probabilities of the Cu core levels were adjusted such that the total of all annihilation probabilities added up to the contribution from Cu atoms to the total Doppler spectra. Annihilation probabilities for various orbitals of the Cu obtained from our fits match reasonably well with the calculated Doppler spectrum for bulk Cu (limiting the $r_s = 2.67\,a.\,u.$) as given in Table 4. The percentages of carbon and sulfur are also mentioned in Table 4.

The ratio of the bulk Cu Doppler spectra to the surface Cu Doppler spectra (Fig. 7 (b)) shows that the surface Doppler spectrum differs from its bulk spectra. Comparison of the surface and bulk annihilation probabilities derived from the fits indicate that the reasons for the difference are (i) enhanced valence annihilations at the surface compared to the bulk and (ii) the significant suppression of the annihilation with 3d electrons at the surface.

### vii. Amorphous Se – as-inserted and sputtered.

We performed PAES and CDBS on one micrometer-thick amorphous Se film on polycrystalline Cu foils. PAES of the as-inserted amorphous Se film (Fig. 8 (a)) show that the surface primarily has oxygen, carbon, and sulfur contaminants. However, the presence of Se can be seen through its $M_{4,5}VV$ Auger peak (marked in Fig. 8(a)). After a minor sputtering of the amorphous Se film, the Auger peaks corresponding to the contaminants are reduced, and the $M_{4,5}VV$ Auger peaks of Se become prominent. The $M_{2,3}VV$ peak from the Cu substrate also starts to come to the fore. The complex variations seen in the elemental composition of the topmost atomic layer with sputtering, as evidenced by the PAES, are reflected in the CDBS of the inserted and sputtered amorphous Se surfaces. The differences are better highlighted in the ratio of the spectra taken with the clean Cu surface (Fig. 8 (b)), once again showing that CDBS can identify minor changes to the chemical composition of the topmost atomic layer with a surface selectivity and sensitivity similar to surface electron spectroscopies.

### 4. Conclusion:

In this paper, we have presented CDBS obtained from the external surfaces of clean, adsorbent, or film-covered surfaces. The results we obtained for external surfaces give confidence that CDBS can be used effectively for hidden surfaces, specifically the internal surfaces of porous materials. The CDBS results were complimented for the first time by highly surface and element-specific PAES measurements, giving confidence in the ability of CDBS to be a top-layer selective technique for elemental quantification, which can be used for both external and internal surfaces. While previous studies have shown the sensitivity of CDBS to surface conditions, there has been no investigation in which the surface was characterized using independent positron-based methods that could independently provide the chemical composition at the site of positron annihilation.

One of the primary results of our study is that the CDBS of surfaces could differentiate between clean and adsorbate or thin film-covered surfaces even when the thin film was a single atomic layer (like graphene). Our measurements showed that the method could differentiate between a copper oxide surface and one where the surface has sub-monolayer sulfur coverage or differentiate between dirty Se film and mostly clean Se films. The present measurements also demonstrated the limited impact Ps formation has on the ability of coincidence Doppler Broadening spectroscopy to quantify the elemental environment of the topmost atomic layer of surfaces of various metals, thin films, and adsorbents.

Another important takeaway from our measurements is that the Doppler spectra from the clean surface of an element may differ substantially from the bulk Doppler spectra. The primary factor contributing to that difference may vary depending on the sample involved. Contribution from Ps annihilation to the surface Doppler spectrum is one of the contributing factors for the difference between bulk and surface spectra. However, their contribution is only about ~ 5-15% in the high momentum region for various surfaces. The other possible reason for the difference is the enhanced annihilation with valence electrons at the surface compared to that in bulk.

Our previous Auger-Doppler coincidence papers showed that Doppler spectra predicted using atomic calculations could predict the shape of the Doppler spectra for individual core levels (3p, 4p, 1s levels).  Assuming that this is true for various electron orbitals, including the d shell, we fit the coincidence Doppler data for multiple surfaces by using atomic calculations for individual orbitals but by allowing the contribution from each of these core and valence levels to vary. We could fit the measured spectra better than what can be obtained by directly using the theoretical Doppler spectra.  The coefficients derived from the fits differ from those predicted by the atomic calculations for d electron systems. In particular, our fits show that the d levels (3d in Cu and 4d in Au) are highly suppressed at the surface. We propose that the d electron annihilation suppression is due to the reduced overlap of the surface-trapped positrons with localized d-electrons. Our results provide strong motivation for fully self-consistent calculation of the Doppler spectra from the surface that calculates the positron wavefunction in the surface image potential well and considers the atomic configuration at the surface. Such a fully self-consistent calculation should be able to reproduce the observed difference between the surface and the bulk Doppler spectra observed for the d electron systems.

The multiparameter fit methodology provided an independent method of determining the percent annihilation with Ps, the valence band, and different core levels. The multiparameter fit to the Doppler spectra that gives the annihilation probability with various orbitals or elements on the surface does not require assumptions regarding the detector response, unlike in the case of PAES, where energy-dependent electron transport efficiencies and electron detector efficiencies must be considered.

The results we obtained for external surfaces give confidence that CDBS can be used effectively for hidden surfaces, specifically the internal surfaces of porous materials. To highlight this point, we have compared the ratio curves of various samples, taking two at a time to show their sensitivity and surface selectivity (Fig. 9). Fig. 9(a) indicates that the Doppler spectrum of clean Cu surface with one atomic layer of carbon can be effectively distinguished from the surface Doppler spectrum of clean Au surface. Fig. 9 (b) shows that a single atomic layer of carbon on Cu (SLG-Cu) leads to distinguishable features in the Doppler spectrum that allow us to distinguish it from the Doppler spectrum of the Cu surface with sub-monolayer sulfur coverage. Fig. 9(c) indicates that the Doppler spectrum from sulfur-segregated Cu surfaces differs from the spectrum of bulk Cu, which has some sulfur contamination. Fig. 9(d) shows that CDBS can distinguish a Cu surface with oxide formation from a Cu surface with sub-monolayer sulfur coverage. Finally, Fig. 9(e) shows that through ratio curve analysis of the Doppler spectrum of a thick film of Se, which still has a contribution from the contaminants (carbon, oxygen, and sulfur) and the Cu substrate, we can distinguish it from a surface (as-inserted Cu substrate with sub-monolayer sulfur coverage) with some standard elemental profile.

Thus, the present study has demonstrated the potential of CDBS as a powerful tool for the elemental profiling of metal surfaces, rivaling the sensitivity achieved by PAES. Since the annihilation gamma signal can penetrate through overlayers and the walls of reaction chambers without much loss of information, CDBS can become an indispensable tool for the characterization of internal surfaces of porous materials where we can obtain the surface-sensitive gamma signals from internal surfaces and reaction chambers – regions that are inaccessible using surface selective electron-based spectroscopies.

**Acknowledgments**

AHW, VAC, and ARK acknowledge the support of NSF Grant No. CHE – 2204230. AHW and ARK acknowledge the support of NSF Grant No. NSF-DMR-1338130 & NSF-DMR1508719. AHW acknowledges the support of the Welch Foundation (Y-1968-20180324). JA acknowledges support from awards DE-SC0020065 and DE-SC0000253485. MR acknowledges support from award FNAL-LDRD-2020-027. The work was prepared in part by LLNL under Contract DE-AC52-07NA27344.

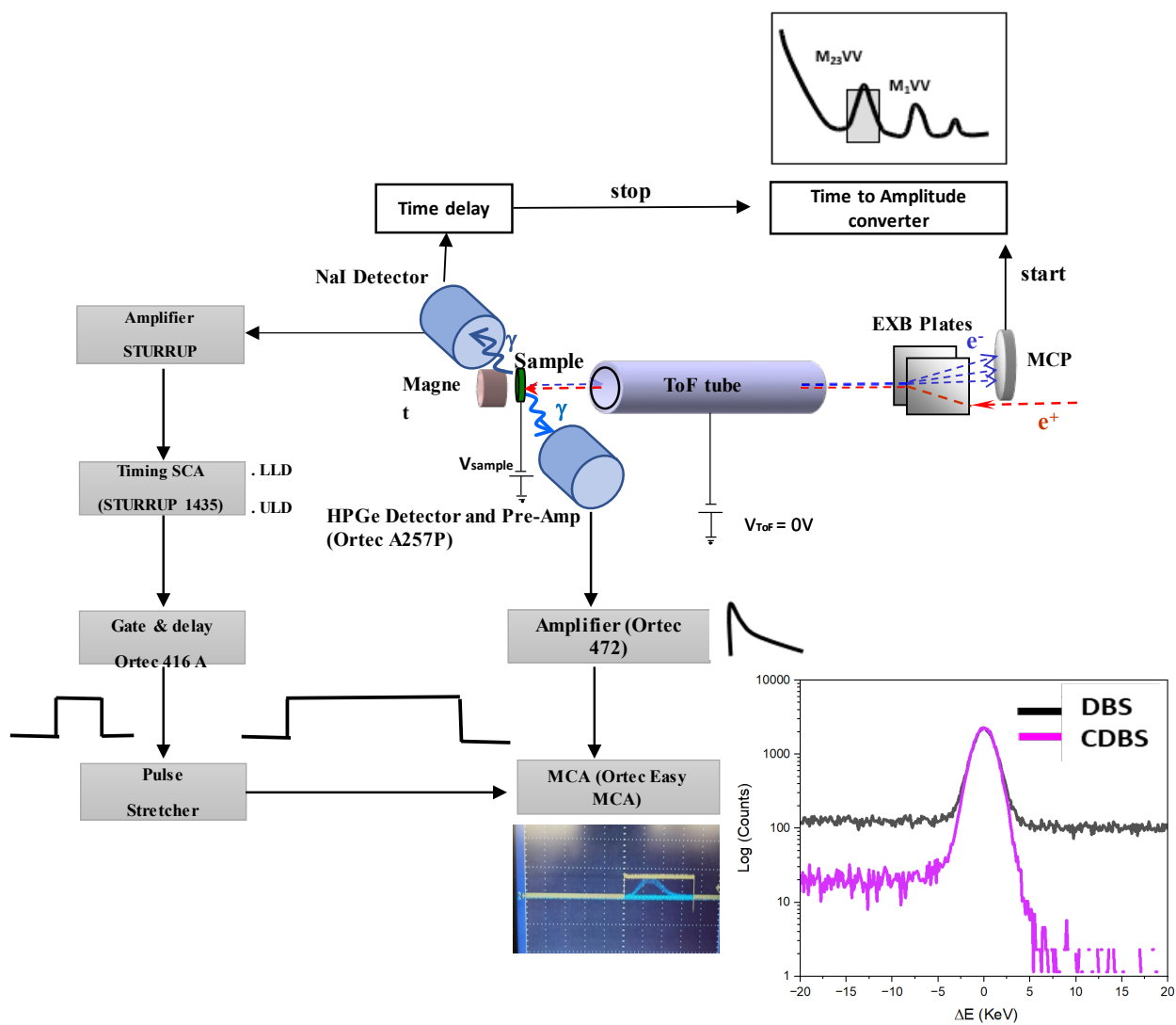

Fig.1.The schematic of the experimental system used to measure CDBS and PAES simultaneously. The system consists of a magnetic bottle time of flight positron-induced Auger electron spectrometer [14, 17]. The Doppler broadened gamma spectra from the annihilation of positrons at the sample are measured using an HPGe detector in coincidence with a NaI (Tl) detector. The NaI (Tl) detector also provides a timing signal for the time-of-flight PAES spectrometer.

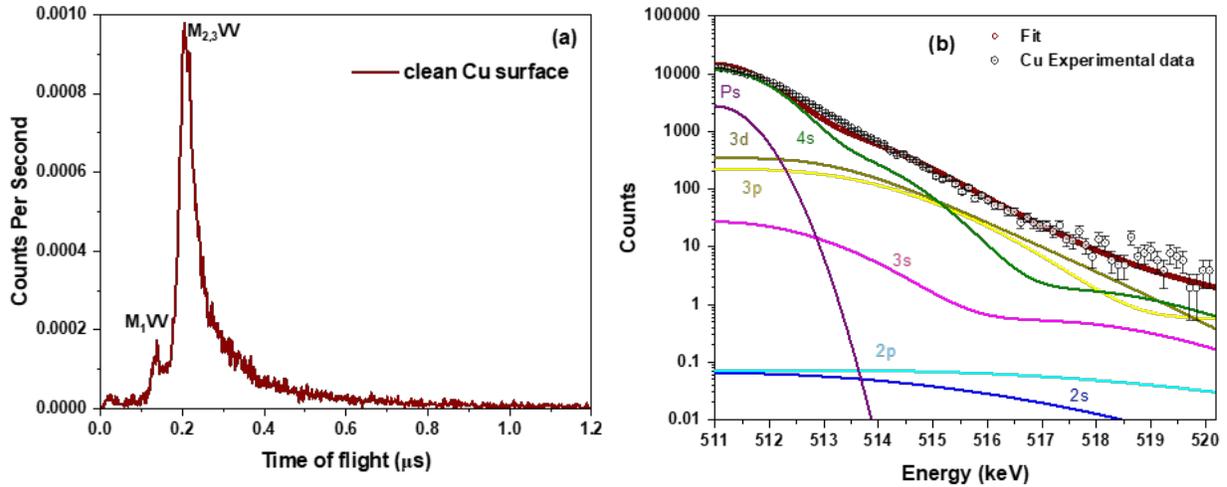

Fig. 2 (a) PAES of clean Cu surface shows the 60 eV M2,3VV and 108 eV M1VV Auger peak. The presence of oxygen or carbon peaks cannot be observed within the detection capability of PAES, confirming a clean Cu surface during the measurements. (b) The annihilation gamma spectrum was measured from a clean Cu surface using DBS. The fit to the measured spectrum using the gamma spectrum obtained from atomic calculations. The contribution of individual core levels of Cu and that from the para positronium annihilation obtained from the fit is shown.

| Orbital | Fit | Atomic Calculation with $3d^{10}4s^1$ | Atomic Calculation with $3d^94s^2$ |
|---|---|---|---|
| 4s | 0.735 | 0.42 | 0.65 |
| 3d | 0.067 | 0.50 | 0.29 |
| 3p | 0.054 | 0.062 | 0.048 |
| 3s | 0.003 | 0.016 | 0.012 |
| 2p | 8.61E-05 | 4.2E-4 | 3.1E-4 |
| 2s | 3.82E-05 | 1.87E-4 | 1.37E-4 |
| 1s | 1.11E-07 | 5.5E-7 | 4.0E-7 |
| p-Ps annihilation | 0.12 | 0 | 0 |
| o-Ps pick-off | 0.03 | 0 | 0 |

**Table 1. Annihilation probability obtained from atomic calculations compared to that obtained from the fit to the measured spectra. Please note the atomic calculation does not include a contribution from Ps formation and annihilation.**

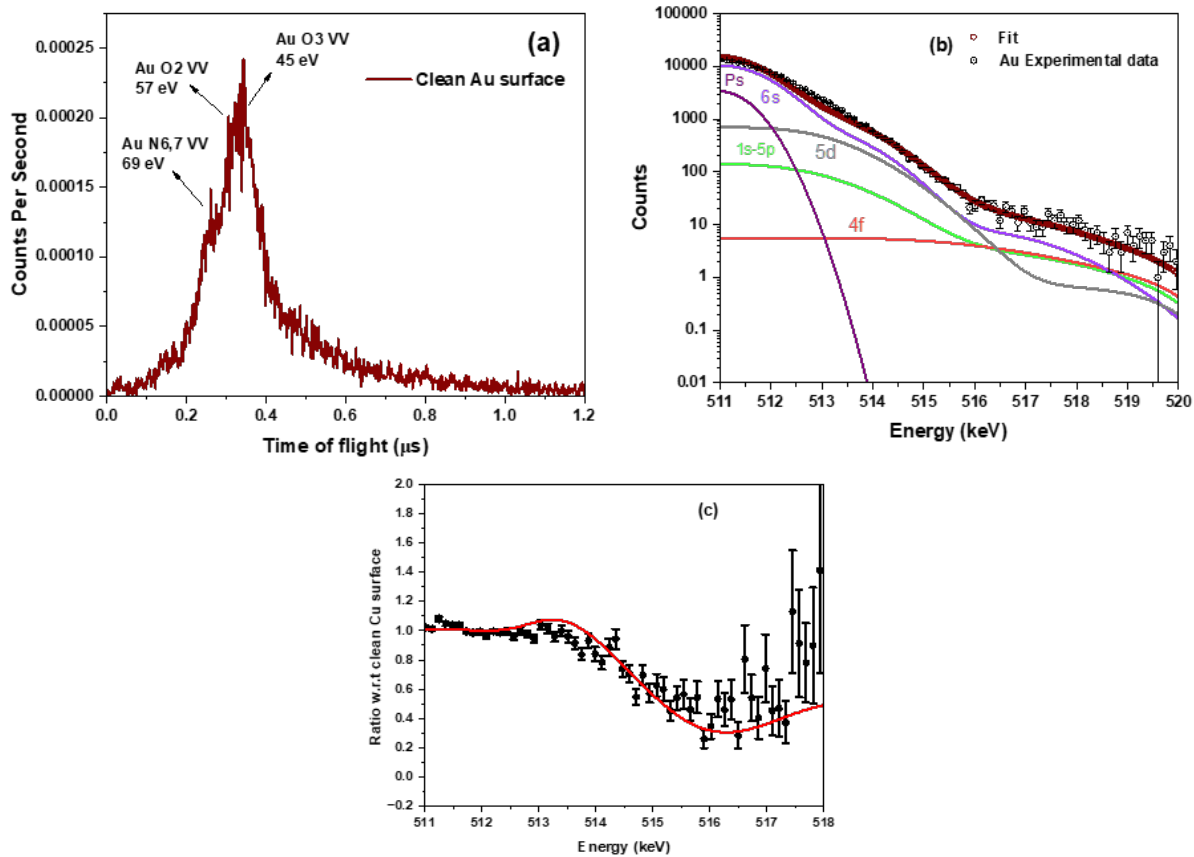

Fig. 3 (a) PAES of clean Au surface shows the N6,7VV, O2VV, and O3VV Auger peaks. The presence of oxygen or carbon peaks cannot be observed within the detection capability of PAES, confirming a clean Au surface during the measurements. (b) Annihilation gamma spectrum from clean Au surface. The fit to the measured spectrum using the gamma spectrum obtained from atomic calculations. The contribution of individual core levels of Au and that from the para positronium annihilation obtained from the fit is shown. (c) The ratio of the measured Doppler spectrum from clean Au surface to the Doppler spectrum from clean Cu. The solid line is the ratio of the fits to the measured Doppler spectra of Au and Cu.

**Table 2. Annihilation probability obtained from atomic calculations compared to that obtained from the fit to the measured spectra. Please note the atomic calculation does not include a contribution from Ps formation and annihilation.**

| Orbital | Fit | Atomic Calculation with $5d^{10}6s^1$ |
|---|---|---|
| 6s | 0.70 | 0.25 |
| 5d | 0.101 | 0.65 |
| 5p | 0.014 | 0.070 |
| 5s | 0.004 | 0.011 |
| 4f | 0.002 | 0.007 |
| 4d | 0.001 | 0.0017 |
| 4p | 2.27E-04 | 5.57E-04 |
| 4s | 4.53E-05 | 1.13E-04 |
| 3d | 1.1E-06 | 2.72E-06 |
| 3p | 1.12E-06 | 2.74E-06 |
| 3s | 3E-7 | 7.20E-07 |
| 2p | 6.4E-10 | 1.60E-09 |
| 2s | 8.9E-10 | 2.23E-09 |
| 1s | 3E-13 | 6.90E-13 |
| p-Ps annihilation | 0.12 | 0 |
| o-Ps pick-off | 0.03 | 0 |

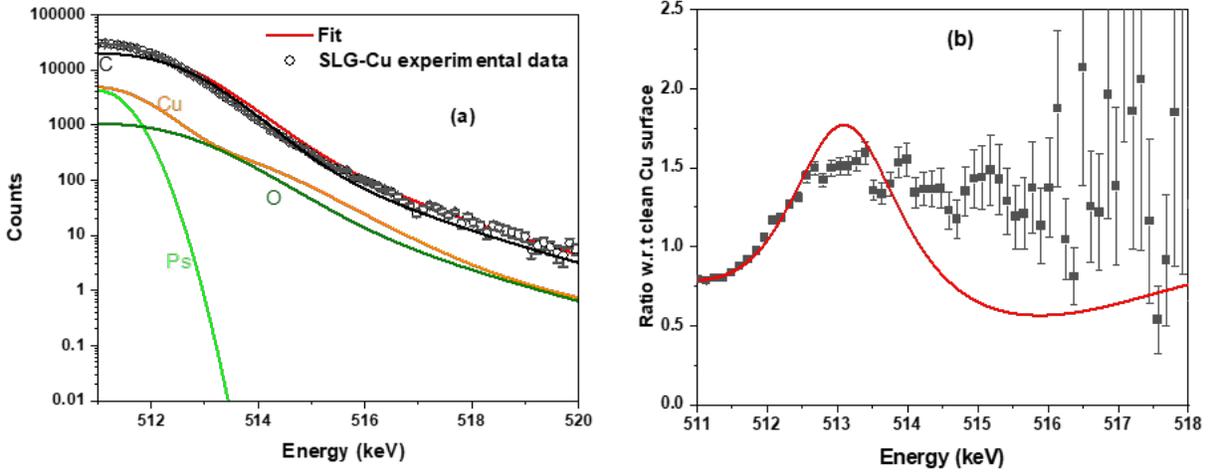

Fig. 4 (a) The annihilation gamma spectrum obtained using DBS of SLG on Cu using a 2 eV positron beam. The fit to the measured spectrum using gamma spectrum obtained from atomic calculations and previous fits to clean Cu spectra. The contribution of individual elements to the total spectra is shown. (b) The ratio of the measured Doppler spectrum from single-layer graphene to the Doppler spectrum from clean Cu. The solid line is the ratio of the fits to the measured Doppler spectra of SLG and Cu.

**Table 3. Annihilation probability obtained from a fit to the measured spectra. Here, the spectra of C and O used for fitting have been obtained directly from the atomic calculations. The spectra for Cu have been obtained from the fit to the clean Cu surface data.**

| Element | Fit |
|---------|-----|
| C | 0.80 |
| Cu | 0.135 |
| O | 0.05 |
| p-Ps annihilation | 0.06 |

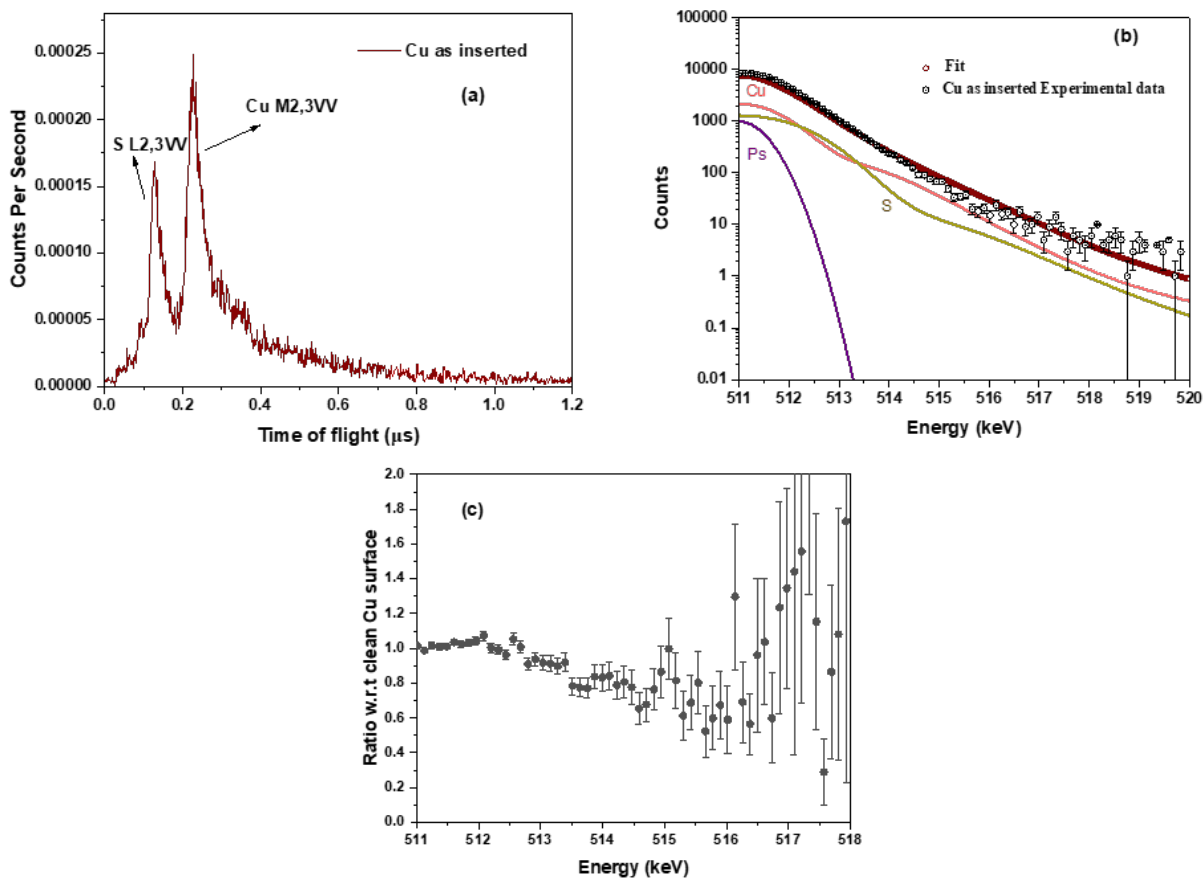

Fig. 5 (a) PAES of as inserted Cu surface, which has sulfur segregation. The spectrum shows the $M_{2,3}VV$ Auger peak of Cu and the $L_{2,3}VV$ Auger peak from sulfur. It is possible to have trace amounts of oxygen or carbon on the surface. However, they are masked by prominent sulfur peaks in the PAES spectrum. In the fit, carbon is included as o-Ps annihilation pickoff, but part of that contribution might be from the surface carbon. (b) The annihilation gamma spectrum measured from as inserted Cu surface. The fit to the measured spectrum using the gamma spectrum obtained from the atomic calculation for sulfur and the fit to the clean Cu data. (c) The ratio of the measured Doppler spectrum from as inserted Cu surface with sulfur surface segregation to the Doppler spectrum from clean Cu.

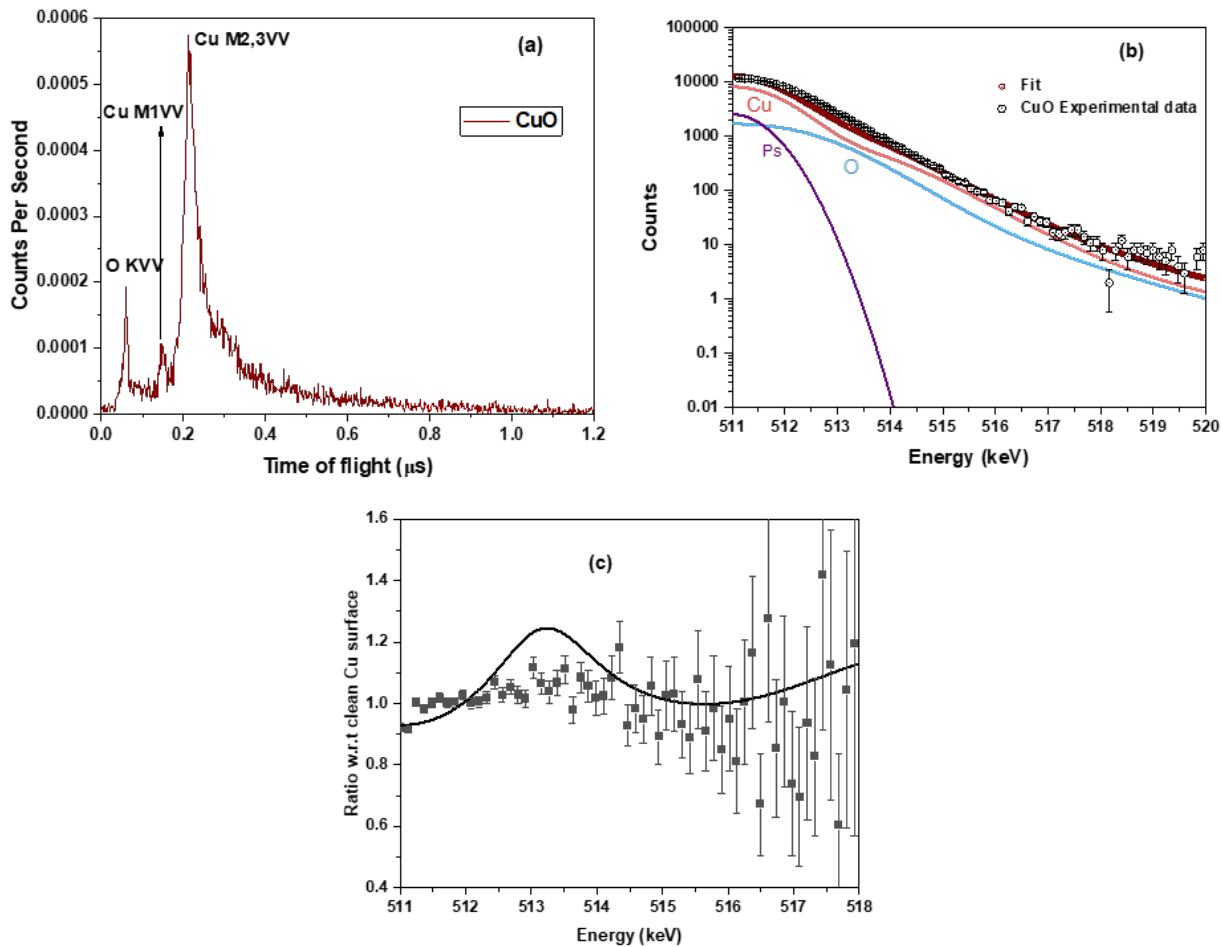

Fig. 6 (a) PAES of oxidized surface of Cu after sputter cleaning shows the M1VV and M2,3VV Auger peaks of Cu and KVV Auger peak of oxygen. Within the detection capability of PAES, the presence of carbon or sulfur peaks cannot be observed, confirming a clean oxidized surface during the measurements. (b) The annihilation gamma spectrum obtained using the DBS of the oxidized Cu surface. The measured spectrum was fitted using the gamma spectrum obtained from the atomic calculation for oxygen and the fit to the clean Cu surface in Fig.2. The contribution of individual elements and that from the para positronium annihilation obtained from the fit is shown. (c) The ratio of the measured Doppler spectrum from oxidized Cu surface to the Doppler spectrum from clean Cu. The solid line is the ratio of the fits to the measured Doppler spectra of oxidized Cu surface and Cu.

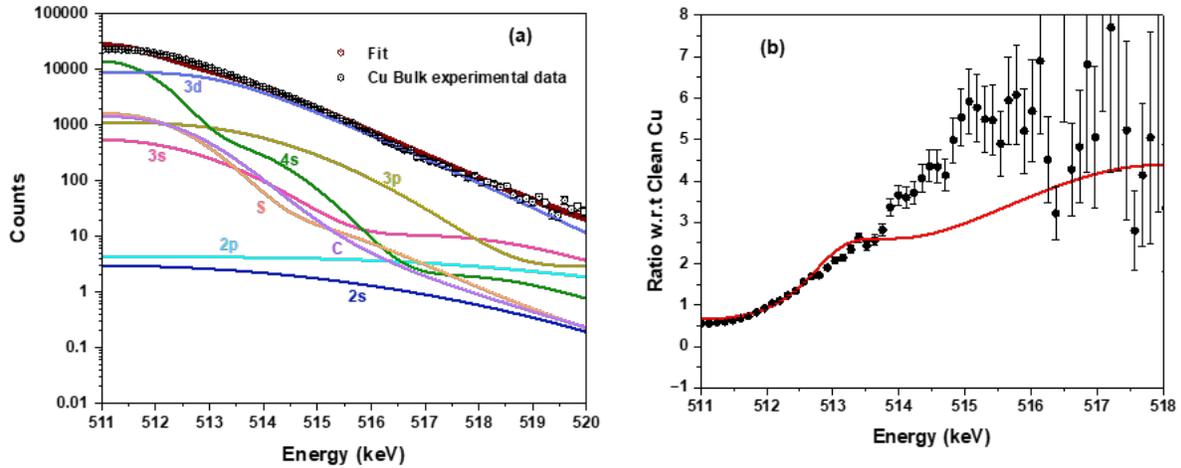

Fig. 7 (a) Doppler broadening spectrum measured from bulk Cu using a 20keV positron beam. The measured spectrum was fit using the gamma spectrum of bulk Cu obtained from the atomic calculations and the calculated Doppler broadened spectrum for sulfur and carbon. Sulfur and carbon were included as common contaminants in pure polycrystalline Cu samples. The contributions of the individual core levels of Cu to the total spectrum were allowed to vary. The contribution of carbon, sulfur, and the individual core levels of Cu obtained from the fit is shown separately. (b) The ratio of the measured Doppler spectrum from bulk Cu to the Doppler spectrum from a clean Cu surface. The solid line is the ratio of the fits to the measured Doppler spectra of bulk Cu and clean Cu surface.

**Table 4. Annihilation probability obtained from atomic calculations compared to that obtained from the fit to the measured spectra.**

| Orbital | Fit | Atomic Calculation with $3d^{10}4s^1$ |
|---------|-----|----------------------------------------|
| 4s | 0.29 | 0.297 |
| 3d | 0.52 | 0.592 |
| 3p | 0.069 | 0.0871 |
| 3s | 0.022 | 0.0220 |
| 2p | 6.25E-04 | 0.00059 |
| 2s | 2.76E-04 | 0.00026 |
| 1s | 8.61E-07 | 7.69E-07 |
| Sulfur | 0.048 | 0 |
| Carbon | 0.048 | 0 |

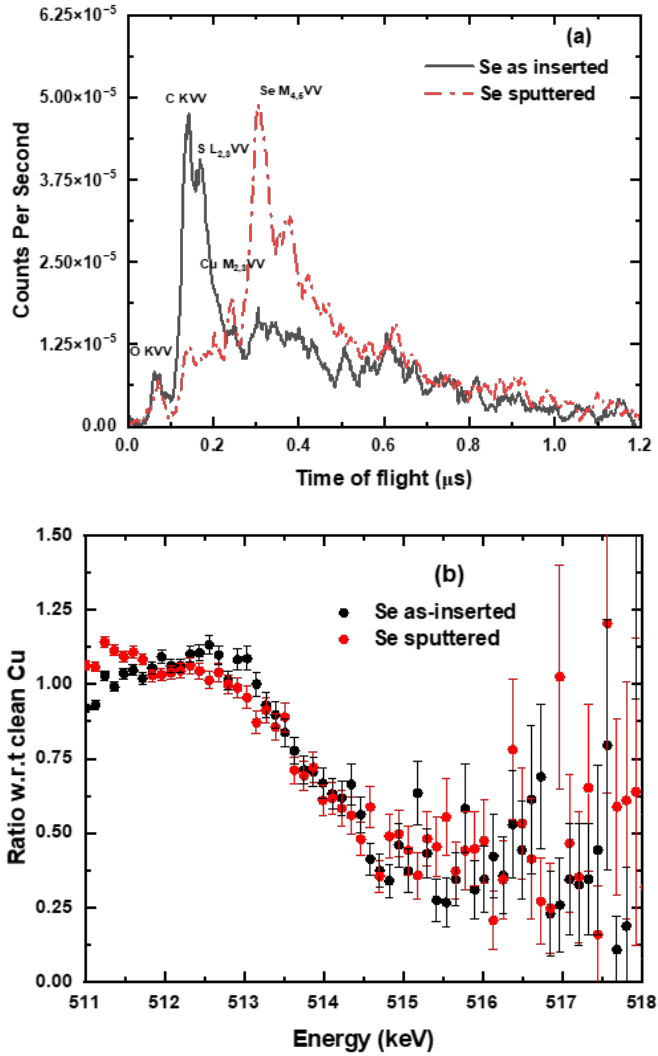

Fig. 8 (a) PAES of as inserted Se film on Cu and the spectrum after minor surface sputtering. The Auger peaks corresponding to the contaminants reduce post-sputtering, and the Auger peak of Se becomes prominent. (b) The ratio of the DBS of the inserted and sputtered Se film to the clean Cu surface shows that the complex variation of elemental composition results in measurable changes in the annihilation gamma spectra.

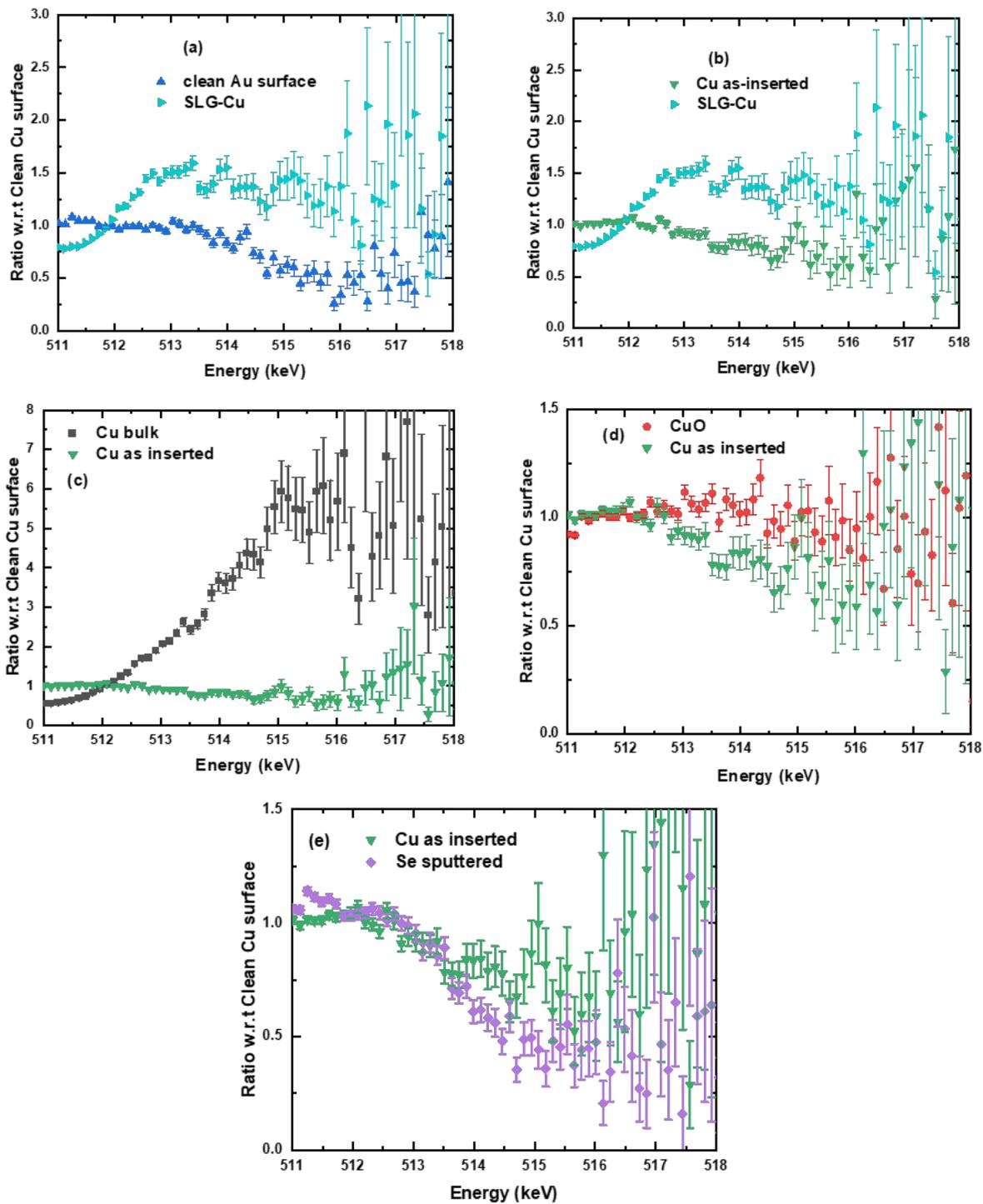

Fig. 9 (a) – (e) Comparison of the ratio of the measured Doppler spectrum from various clean, adsorbent, or thin film-covered surfaces to the Doppler spectrum from a clean Cu surface demonstrating the ability of CDBS to be a top layer selective technique to assay the elemental composition of external or more importantly internal surfaces.